\documentclass[aip,reprint,
amsmath,amssymb,amsfonts,superscriptaddress,twocolumn, footinbib]{revtex4-2}
\usepackage[german,american,english]{babel}
\usepackage{graphicx}
\usepackage{graphics}
\usepackage{dcolumn}
\usepackage{multirow}
\usepackage{bm}
\usepackage{latexsym}
\usepackage{layout}
\usepackage{verbatim}
\usepackage{epsfig}
\usepackage{graphicx}
\usepackage{amsbsy}
\usepackage{xcolor}
\usepackage[T1]{fontenc}
\usepackage[caption=false]{subfig}
\newcommand{\bea}{\begin{eqnarray*}}
	\newcommand{\eea}{\end{eqnarray*}}
\newcommand{\bne}{\begin{equation*}}
\newcommand{\ede}{\end{equation*}}

\newcommand{\bnen}{\begin{equation}}
\newcommand{\eden}{\end{equation}}
\newcommand{\bean}{\begin{eqnarray}}
\newcommand{\eean}{\end{eqnarray}}
\newcommand{\bsen}{\begin{subequations}}
	\newcommand{\esen}{\end{subequations}}
\newcommand{\ba}{\arraycolsep 0.3ex \begin{array}{rl}}
\newcommand{\ea}{\end{array}}
\newcommand{\bna}{\begin{array}}
	\newcommand{\eda}{\end{array}}
\newcommand{\bnm}{\begin{enumerate}}
	\newcommand{\edm}{\end{enumerate}}

\def\pz{{\partial}}

\def\Bk{{\bm k}}

\def\BE{{\bm E}}

\def\CR{{\mathcal R}}
\def\BCR{{\bm\CR}}

\def\frac#1#2{{\textstyle{#1 \over #2}}}

\def\Tra{\mathop{\textsf{Tr}}}

\def\ns{^{\vphantom{*}}}

\def\half{\frac{1}{2}}

\begin{document}
	
\title{Ge as an orbitronic platform: giant in-plane orbital magneto-electric effect in a 2-dimensional hole gas}

\author{James H. Cullen}
\altaffiliation{\textbf{Author to whom correspondence should be addressed:} james.cullen@unsw.edu.au}
\affiliation{School of Physics, The University of New South Wales, Sydney 2052, Australia}

\author{Dimitrie Culcer}
\affiliation{School of Physics, The University of New South Wales, Sydney 2052, Australia}

\begin{abstract}
Increasing demand for computational power has initiated the hunt for energy efficient and stable memory devices. This is the overarching motivation behind the recent rise of \textit{orbitronics}, which looks to harness the orbital angular momentum of charge carriers in computing devices. Orbitronic devices require materials with efficient generation of orbital angular momentum (OAM). In 2D materials, OAM can be electrically generated via the orbital magneto-electric effect (OME). In this paper we report the calculation of the OME in 2 dimensional hole gases (2DHGs). We show that the OME in Ge holes is very large, for an applied electric field of the order $10^4$ V$/$m the OAM density is of the order $10^{12}$ $\hbar/$cm$^{2}$. { Furthermore, we find the OME to be an order of magnitude larger than the Rashba-Edelstein effect in 2DHGs. The OME we calculated in 2DHGs generates OAM aligned in the plane and arises due to transitions between heavy and light hole states}
, which is unique to this system. Our results put Ge, as well as other p-type semiconductors, forward as strong candidates for building future orbitronic devices.
\end{abstract} 

\date{\today}
\maketitle

\section{Introduction}

Recent years have seen a surge of interest in the generation of orbital angular momentum (OAM) and its underlying mechanisms \cite{Exp-OHE-Ti-Nat-2023-Hyun-Woo,Orbitronics-in-action, Rhonald-Rev, OC-Rev-EL-2021-Yuriy, wang2024orbitronics}. The motivation for studying orbital effects stems from their application in \textit{orbitronic} devices, which use charge carrier orbital angular momentum and its interaction with magnetic degrees of freedom for the storage and manipulation of information. The electrical generation and transport of Bloch electrons' OAM has witnessed significant experimental progress \cite{OT-FM-PRB-2021-YoshiChika, OT-OEE-NatComm-2018-Haibo,OT-PRR-2020-Hyun-Woo,OT-NatComm-2021-Kyung-Jin, PhysRevB.106.184406, Exp-OT-PRR-2020,Exp-OT-CommP-2021-Byong-Guk, LS-conversion-CP-2021-Byong-Guk,OOS-Cvert-2020-PRL-Mathias, OHE-Hetero-PRR-2022-Pietro, OHE-OT-large, L-S-OT-2023,OHE-Binghai, tokura2019magnetic, go2023long, Niels_Orbitalsplitter, 10.1063/5.0106988,el2023observation,Exp-OEE-PRL-2022-Jinbo, santos2024negative}. The interaction between orbital angular momentum and an adjacent magnetisation is referred to as \textit{orbital torque}, that is, a torque exerted on a magnetisation by a non-equilibrium OAM density. This is the orbital analogue of the spin torque that has received considerable attention in magnetic systems \cite{sakai2014, yasuda2017current, tatara2007spin, kohno2006microscopic, belashchenko2019first, CI-SOT-RMP-2019-Manchon, nikolic2020first, gambardella2011current, PhysRevB.92.014402, PhysRevB.75.214420, brataas2012current, PhysRevB.88.085423, PhysRevB.91.214401}. There are two primary mechanisms for the electrical generation of OAM; (i) the electrical generation of orbital currents, known as the orbital Hall effect (OHE) \cite{Orbitronics-PRL-2005-Shoucheng,ISHE-IOHE-PRB-2008-Inoue,OHE-PRL-2009-Inoue, OHE-metal-PRM-2022-Oppeneer, OHE-BiTMD-PRL-2021-Tatiana,canonico2020two, PhysRevLett.132.106301-Giovanni, veneri2024extrinsic}. (ii) the electrical generation of orbital densities, known as the orbital magneto-electric effect (OME) \cite{thonhauser2011review, OM-metal-PRB-2021-Xiaocong, malashevich2010theory, PhysRevB.102.184404, osumi2021kinetic, OEE-NL-2018-Shuichi, Titov_EdgeOM, lee2024orbital, PhysRevB.107.094106}. In 2D materials the OME is the most sensible avenue for the generation of orbital torques. This is because the OHE can only flow in-plane, only leading to an edge accumulation, whereas the OME will generate an orbital polarisation throughout the sample. As such, the OME is the most promising avenue for generating large orbital torques in 2D materials, as recently reported in strained twisted bilayer graphene \cite{he2020giant}. 

A primary question in the field of orbitronics is concerned with identifying mechanisms and materials that maximise OAM generation; here we focus on 2D materials and the OME. In experiment, orbital and spin effects cannot be definitively distinguished, and for now the primary indication of their relative magnitudes comes from theoretical calculations. We recently showed that a giant OHE is present in Ge \cite{cullen2025ge}, putting forward $p$-type semiconductors as a potential platform for orbitronic devices. The OME has been studied using the modern theory of orbital magnetisation in both clean systems \cite{malashevich2010theory, OM-metal-PRB-2021-Xiaocong, cullen2025quantum}, and in systems with disorder scattering and transport relaxation \cite{OEE-scalar-potential, OEE-NL-2018-Shuichi, osumi2021kinetic}. The modern theory of orbital magnetisation shows that orbital magnetisation of charge carriers described by Bloch wavefunctions can be constructed via the Berry connection. This theory was derived using both semiclassical and Wannier approaches \cite{Theroy-OM-PRL-2007-Qian, Resta-PhysRevLett.95.137205, Vanderbilt_2018, Resta-PhysRevResearch.2.023139, Resta-PhysRevB.74.024408, thonhauser2011review}, and has been used to describe the orbital magnetisation both in and out of equilibrium \cite{thonhauser2011review, OM-metal-PRB-2021-Xiaocong, malashevich2010theory, cullen2025quantum}. The OAM considered in the modern theory differs from the orbital effects considered using the atom-centred approximation (ACA), in which the OAM of atomic orbitals is considered. The connection between these two approaches to the calculation of orbital effects is still ambiguous and the results generated by each approach can differ by orders of magnitude \cite{IOHE-XIV-PRB-2021-Hyun-Woo,cullen2025ge}. In the ACA orbital densities and currents look almost identical to their spin analogues but with the spin operator replaced by the OAM of the atomic orbitals corresponding to each band. In the modern theory the OAM operator takes the form of $m(\bm r\times\bm v)$ where $\bm r$ and $\bm v$ are the position and velocity operators, suitably antisymmetrised. Within this theory the OAM can be separated in local circulation and itinerant circulation, describing orbital motion within the unit cell and within the lattice respectively \cite{malashevich2010theory, Vanderbilt_2018, thonhauser2011review, cullen2025quantum}. { However, the atom-centred approximation is assumed to be subsumed by the modern theory as it accounts for all forms of OAM \cite{OHE-PRB-2022-Manchon}, an ab-initio calculation of the orbital magnetisation showed that the modern theory recreated the atom-centred approximation results for materials in which the OAM is localised about the atomic centre \cite{OM-Berry-PRB-2016-Mokrousov}.}

In this work we consider electrically generated orbital densities in 2-dimensional hole gases (2DHGs) within the modern theory of orbital magnetisation. Our investigation of the OME in 2DHGs is motivated in part by a recent study that demonstrated large torque generated by a 2DHG in a hydrogen-terminated diamond system \cite{SOT2DHGDiamond} {as well as observations of the OHE in Si and Ge\cite{shirai2025signature, santos2024negative, matsumoto2025observation}}. 
We find that there is an OME in 2DHGs with broken inversion symmetry, {the OAM generated is itinerant and as such is not captured by the ACA}. 
Due to spin-orbit coupling effects holes in semiconductors are described by a $4\times4$ Hamiltonian with total angular momentum $J=3/2$. The good quantum number for holes in the bulk is the total angular momentum $J$ projected onto the wavevector $\bm k$\cite{chow&koch}, holes with this projection equal to $\pm3/2$ are known as heavy holes while those with $\pm1/2$ are known as light holes due to their differing effective masses. For a 2DHG with an out-of-plane confinement the Hamiltonian is obtained by treating the out-of-plane component of the crystal momentum as an operator and projecting onto the out-of-plane wavefunctions\cite{chow&koch}. We find the OME in 2DHGs to be in-plane, which is in sharp contrast to other known 2D systems \cite{OM-PRB-2021-Xiaocong, OM-metal-PRB-2021-Xiaocong, OEE-NL-2018-Shuichi, osumi2021kinetic, PhysRevB.107.094106, cullen2025quantum}. This in-plane OAM comes from the matrix elements of the position and velocity operators connecting heavy hole and light hole states. If the confinement potential is asymmetric the heavy and light holes will have different centres of mass in the out-of-plane direction, and transitions between these states allow for movement in the out-of-plane direction. Hence, 2DHGs can exhibit an in-plane OAM. 

The OME calculated in this work arises from the electrically induced shift in the Fermi contour, which is proportional to the transport relaxation time. Our numerical estimates of the OME in Ge 2DHGs are very large due to their high mobilities and transport relaxation times. We estimate that the OME in Ge 2DHGs exceeds that of the Rashba-Edelstein effect (REE) in 3D topological insulators by up to two orders of magnitude. This is largely due to the excellent sample quality of p-type Ge in which hole mobilities of the order order $10^6$ cm$^2/$Vs have been demonstrated \cite{stehouwer2023germanium, Borsoi2025nature}, while TI surface state mobilities are generally of the order $10^3$-$10^4$ cm$^2/$Vs \cite{gehring2012growth, koirala2015record, shrestha2017extremely, wang2017observation}. We compare our results for the OME in Ge with the REE in 3D topological insulators, as these materials are known to produce very efficient spin torques \cite{SOT-TM-Rev}. As such, we propose Ge as a promising orbitronic material. In particular, Ge is theoretically predicted to have a giant OHE \cite{cullen2025ge}. Additionally, the proximity to Si microfabrication ensures high sample quality, a property that has made Ge a material of choice for semiconductor quantum computing \cite{scappucci2021germanium, fang2023recent}. Altogether these observations make Ge an ideal platform for the realisation of energy efficient orbitronic devices.

The outline of this paper is as follows: In Sec. II we outline our methodology, including the Kohn-Luttinger Hamiltonian, transport formalism based on the Liouville equation, and approach to determining the non-equilibrium density matrix together with the OAM density. In Sec. III we present our results for the OME in 2DHGs, showing how they vary for different confinements and gate voltages. In Sec. IV we discuss the implications of our findings for orbitronics as well as for the modern theory of orbital magnetisation. We also compare our results in Ge with the surface states of topological insulators. We end with a summary and outlook. 

\begin{figure}[tbp]
\begin{center}
\includegraphics[trim=0cm 0cm 0cm 0cm, clip, width=\columnwidth]{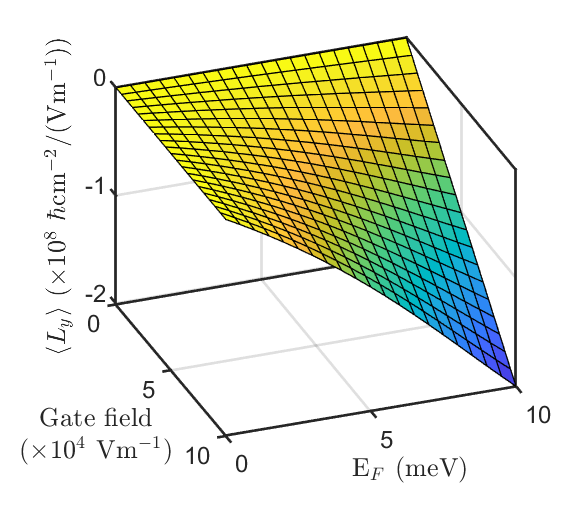}
\caption{\label{Fig:Ge_OAM3D}The OAM density per unit field in a Ge 2DHG for a confinement of 20 nm and relaxation time $100$ ps for various Fermi energies and gate fields.}
\end{center}
\end{figure}

\section{Methodology}
We consider a system described by the Hamiltonian $H = H_b +V_0(z)+U+H_{\bm E}$, where $H_0$ is the band Hamiltonian, $V_0$ is the confinement potential, $U$ is the disorder potential and $H_{\bm E}$ is the electric potential. The electric potential $H_{\bm E}$ consists of a gate field in the $z$-direction $-eFz$ and another applied field in the $x$-direction $-eEx$. We use the Luttinger Hamiltonian confined in the spherical approximation $\gamma_2\approx\gamma_3$, where we use $\bar{\gamma}=(\gamma_2+\gamma_3)/2$. The band Hamiltonian is \cite{luttinger1955motion}
\begin{equation}
    H_b = 
    \begin{pmatrix}
        P+Q & 0 & L & M\\
        0 & P+Q & M^* &-L^*\\
        L^* & M & P-Q & 0\\
        M^* & -L & 0 & P-Q
    \end{pmatrix}\,,
\end{equation}
where, in the spherical approximation, $P=\alpha\gamma_1 (k_{\parallel}^2+k_z^2)$, $Q=\alpha\bar{\gamma} (k_{\parallel}^2-2k_z^2)$, $L=-2\sqrt{3}\alpha\bar{\gamma}(k_x - ik_y)k_z$, $M=-\sqrt{3}\alpha\bar{\gamma}(k_x-ik_y)^2$ and $\alpha=-\hbar^2/2m$. Note, since we are considering the case in which the holes are confined in the $z$-direction, we treat $k_z$ as an operator $k_z=-i\partial/\partial z$ \cite{chow&koch}. We consider a system where the confining potential $V_0$ is an infinite square well with a width of $d$. To obtain the $z$ component of the eigenstates we consider only the diagonal elements of the Hamiltonian, the variational solution for the ground state $z$ wavefunction for this system is the Bastard wavefunction \cite{bastard1983variational}. For our calculation we only include the ground state wavefunctions for the heavy and light holes and work with a $4\times 4$ Hamiltonian.

The OAM operator is defined as $\bm L = m (\bm r \times \bm v-\bm v\times\bm r)$. To calculate the expectation value of the OAM operator we take the trace of the operator with the density matrix $\Tra L_a \rho$. If the centre of mass of the heavy and light holes are separated in the $z$-direction, all components of the OAM $\bm L$ can exist. The $z$ operator is straightforward to calculate since the holes are confined by a potential well in this direction, whereas the position operator in the $x$-$y$ plane is difficult to treat since the states are delocalised. 
{We use the formalism introduced in Refs.~\onlinecite{Hong-PSHE-PRB, liu2025quantumOHE, cullen2025quantum} to calculate the matrix elements of the combination of the density matrix $\rho$ with $x$ and $y$ operators. This formalism involves employing the effective displacement $\bm \Xi = 1/2\{\bm r,\rho\}$, which can be calculated via the quantum Liouville equation.} 
To evaluate the expectation value of all components of the OAM we calculate
\begin{align}
    \langle L_x\rangle =& -m\Tra(v_z\Xi_y-\half\{z,v_y\}\rho)\,,\\
    \langle L_y\rangle =& -m\Tra(\half\{z,v_x\}\rho-v_z\Xi_x)\,,\\
    \langle L_z\rangle =& -m\Tra(v_y\Xi_x-v_x\Xi_y)\,,
\end{align}
where $m$ is the bare electron mass, we have changed the sign of the OAM since we are looking at the OAM of holes. Note, all multiplications of position and velocity operators have been anti-symmetrised.

To calculate electrically induced orbital densities, we need the the nonequilibrium correction to the density matrix in an electric field as well as the effective displacement. We calculate the nonequilibrium correction to the density matrix using the following equation\cite{Interband-Coherence-PRB-2017-Dimi}
\begin{equation}
{\pz\rho\ns_\BE\over\pz t}+{i\over\hbar}\big[H_0\,,\,\rho\ns_\BE\big]+J(\rho_{\bm E})=
-{i\over\hbar}\big[H_{E,x},\rho_0\big],
\label{eq:qke}
\end{equation}
where $H_0=H_b +V_0(z)-eFz$ and $J(\rho)$ is the scattering term. The solution to this equation can be separated into two components $\rho_{\bm E}=n_{\bm E} + S_{\bm E}$, where $n_{\bm E}$ is the band diagonal component and $S_{\bm E}$ is the band off-diagonal component. In this work we use a simple relaxation time approximation {in the weak scattering limit} $J(\rho_{\bm E})=n_{\bm E}/\tau$. In this simple approximation $n_{\bm E}$ captures the extrinsic part of the nonequilibrium density matrix and $S_{\bm E}$ the intrinsic part. {We expect the weak scattering limit to properly capture this system due to the high mobilities in Ge 2DHGs}. The effective displacement ${\bm \Xi}_{\bm E}$ is found from the equation \cite{Hong-PSHE-PRB,liu2025quantumOHE,cullen2025quantum}
\begin{equation}
{\pz{\bm \Xi}\ns_\BE\over\pz t} + {i\over\hbar}\big[H\ns_0\,,\,{\bm \Xi}\ns_\BE\big] =
-{i\over\hbar}\big[H\ns_\BE\,,\,{\bm \Xi}\ns_0\big]
- {1\over 2\hbar}\big\{ D\ns_\Bk H\ns_0\,,\,\rho\ns_\BE\big\}\quad,
\label{eq:qkeXi}
\end{equation}
where $D\ns_\Bk O = \pz\ns_\Bk O - i[\BCR,O]$. The parts of the expectation value of $L_a$ involving the effective displacement are calculated along the same lines as Ref.~\onlinecite{cullen2025quantum}, the remaining terms involving the $z$ operator are simply calculated by finding the matrix elements of $L_a$ and tracing it with the solution to (\ref{eq:qke}), further details of this calculation can be found in the supplementary material.

\section{Results}

\begin{figure}[tbp]
\begin{center}
\includegraphics[trim=0cm 0cm 0cm 0cm, clip, width=\columnwidth]{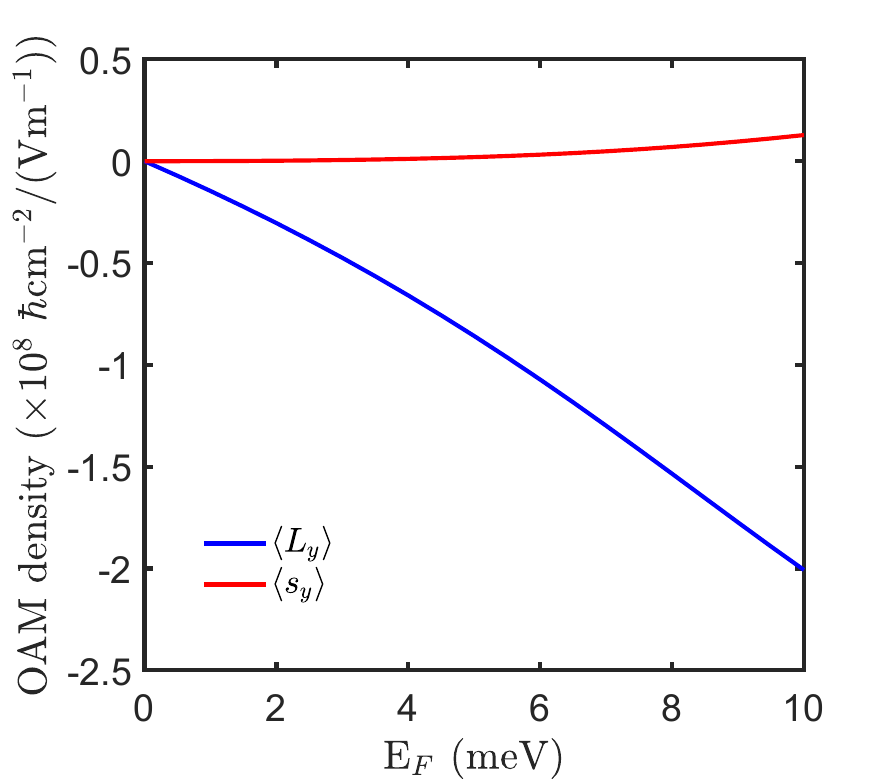}
\caption{\label{Fig:GE_OAM_Sy}The OAM and spin densities per unit field in a Ge 2DHG for a confinement of 20 nm, relaxation time $100$ ps and a gate field of $F=10^5$ V$/$m.}
\end{center}
\end{figure}

We calculated all components of the nonequilibrium expectation value of $\langle \bm L\rangle$ to first order in the electric field. We find all components other than $\langle L_y\rangle$ to be zero. The electrically induced OAM density $\langle L_y\rangle$ is purely extrinsic due to disorder scattering and the intrinsic components are zero, this is shown in the supplementary material. Now, for the model considered the diagonal elements of the $y$ component of the OAM operator can be expressed as
\begin{equation}
    L_y^{mm}=\Delta z C^{mm}\,,
\end{equation}
where $\Delta z=\langle z\rangle_{HH}-\langle z\rangle_{LH}$ is the difference between the $z$ expectation value of the heavy hole and light hole $z$ wavefunctions and the diagonal elements of $C$ are linear in $k_x$. So, the extrinsic contribution to the electrically induced OAM density is calculated as
\begin{equation}\label{Eq:Ly_exp}
    \langle L_y \rangle = \Tra L_y\, n_{\bm E} =  \Delta z \Tra C\, n_{\bm E}\,,
\end{equation}
where $n_{\bm E}$ is linear in the relaxation time $\tau$. For an electric field $\bm E\parallel \hat{x}$, $n_{\bm E}$ will be linear in $k_x$ and (\ref{Eq:Ly_exp}) will be nonzero. The expressions for the OAM include terms that depend on the origin, where $z=0$, these terms can be expressed as the average centre of mass of the heavy and light holes $\Tilde{z}=(\langle z\rangle_{HH}+\langle z\rangle_{LH})/2$. The average centre of mass appears in the diagonal elements of the $z$ operator, we set these terms to zero ($\Tilde{z}=0$) to avoid anything depending on the choice of origin.

The expression for the expectation value of $\langle L_y \rangle$ in (\ref{Eq:Ly_exp}) is proportional to the separation of the centre of masses of the heavy and light holes. Hence, the effect requires inversion symmetry breaking either via a nonzero gate field or due to a asymmetric potential well. Furthermore, we find that the OAM density is linear in the relaxation time $\tau$, {as the intrinsic contribution is zero and the only nonzero contribution is (\ref{Eq:Ly_exp}) which is explicitly linear in the relaxation time}. 
As such, the electrically induced OAM density we have calculated is analogous to the Rashba-Edelstein effect (REE) in a 2D system with broken inversion symmetry, similar phenomenon in other systems have been referred to by the orbital Edelstein effect (OEE) \cite{OEE-NL-2018-Shuichi, Edelstein-PhysRevResearch.3.013275, OEE-scalar-potential, johansson2024theory, PhysRevB.110.184503, lee2024orbital}. The OEE and OME both refer to the same effect; the generation of an OAM density by an applied electric field. As such, in this work we will refer to the effect only as the orbital magneto-electric effect. Note, we found some contributions to the OME that depend on the choice of origin, we chose to ignore these terms due to their arbitrary nature.

\begin{figure}[tbp]
\begin{center}
\includegraphics[trim=0cm 0cm 0cm 0cm, clip, width=\columnwidth]{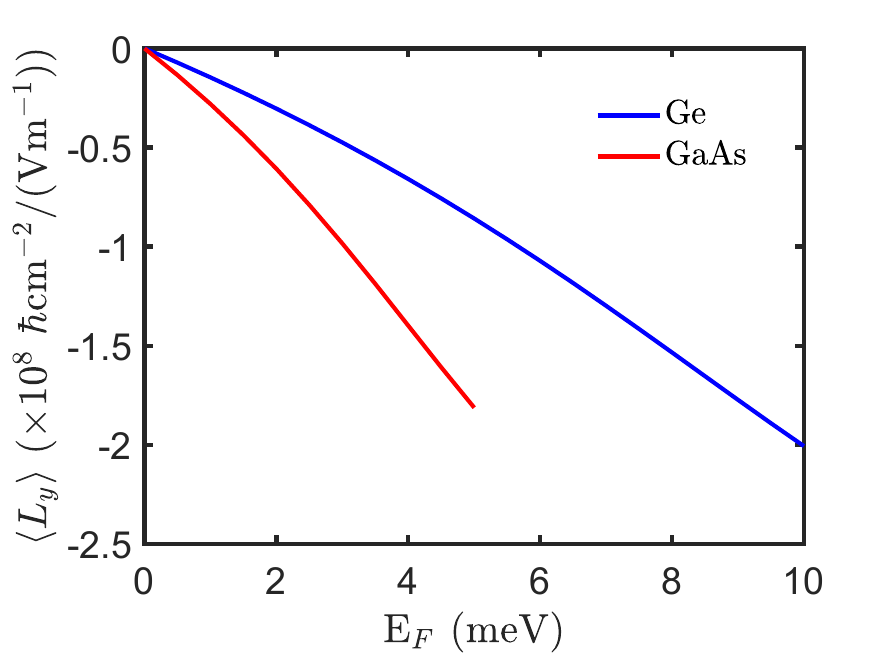}
\caption{\label{Fig:Ge_GaAs}The OAM densities per unit field for Ge and GaAs for a confinement of 20 nm. The OAM densities are plotted for Fermi energies from the bottom of the heavy hole band {to the energy of the first excited state}. Here we use a relaxation time of $100$ ps and a gate field of $F=10^5$ Vm$^{-1}$ for both materials.}
\end{center}
\end{figure}

The OME in a Ge 2DHG is very large, as is shown in Figs.~\ref{Fig:Ge_OAM3D} \& \ref{Fig:GE_OAM_Sy}. For an applied electric field of the order $10^4$ V$/$m the OAM density can be of the order $10^{12}$ $\hbar/$cm$^{2}$. { For a well width of 20 nm, a gate field of $10^5$ V$/$m at a Fermi energy of 10 meV, an applied field of $10^4$ V$/$m will yield an OME equivalent to having a REE effect in which there is $\sim100\%$ spin polarisation in all charge carriers}. We have also included some estimates for the OME in GaAs, shown in Fig.~\ref{Fig:Ge_GaAs}, in which we find the magnitude of the OME to be similar in both Ge and GaAs. For our numerical estimates, we choose a very large relaxation time $100$ ps. We choose a larger relaxation time because Ge and has ultra high mobilities of order $10^6$ cm$^2/$Vs \cite{GaAsholemobility, stehouwer2023germanium, Borsoi2025nature}, indicating that their relaxation times can of the order $100$ ps. Here we have used a simple relaxation time approximation, {we expect there to be some corrections with a proper treatment of disorder including vertex corrections and spin-orbit scattering effects. However, for this work this approximation is sufficient, we are primarily interested in the magnitude of the OME in 2DHGs which is linearly dependent on the relaxation time, that we determine from experimental results. For both a relaxation time approximation and for a proper treatment of disorder, the disorder is generally captured with one parameter; the relaxation time or the impurity density. In either case the estimates for this parameter are determined by comparison of current density calculations, determined in the weak scattering limit by $n_E$, with experimental measurements. Since $n_E$ is also the component of the density matrix responsible for the OME, we do not expect a proper treatment of disorder to substantially modify the magnitude of the OME. Furthermore, we do not expect higher order and spin-orbit scattering corrections to significantly change the magnitude of the OME because of the enormous relaxation times considered, as these corrections usually appear at the $\tau^0$ order \cite{JE-PRR-Rhonald-2022}.} 

\begin{figure}[tbp]
\begin{center}
\includegraphics[trim=0cm 0cm 0cm 0cm, clip, width=\columnwidth]{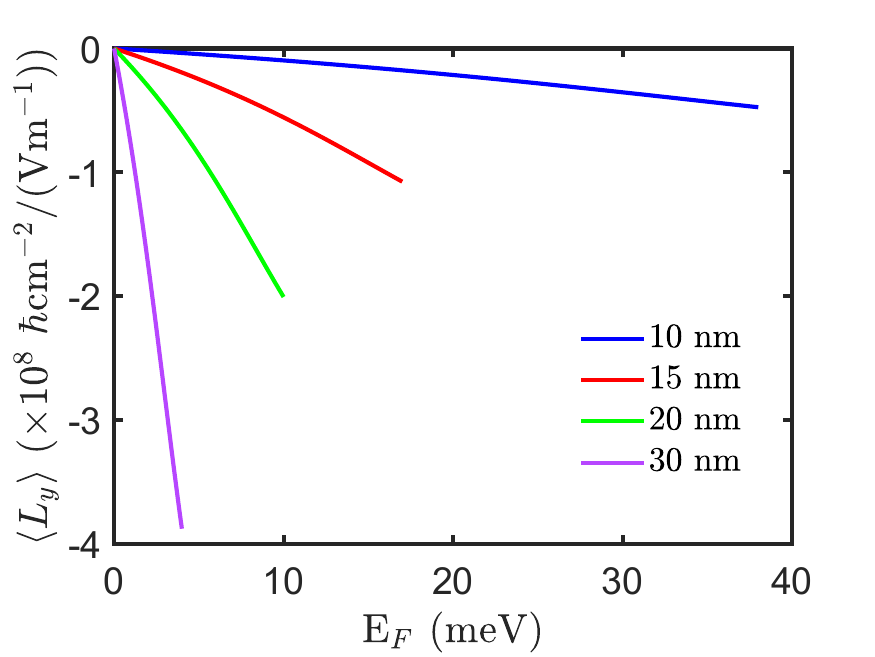}
\caption{\label{Fig:well_width}The OAM densities per unit field for Ge for various well widths. The OAM densities are plotted for Fermi energies from the bottom of the heavy hole band to {the energy of the first excited state.} The relaxation time is $100$ ps and the gate field is $F=10^5$ Vm$^{-1}$.}
\end{center}
\end{figure}

We find that the OME in 2DHGs is an order of magnitude larger than the REE, shown in Fig.~\ref{Fig:GE_OAM_Sy}. This is consistent with our previous calculations in which orbital effects dominate spin effects \cite{cullen2025ge, cullen2025giant}. We find that our REE results are in reasonable agreement with Ref.~\onlinecite{REE2DHG} in the parameter regime in which their effective Hamiltonian is applicable. For our calculation of the REE in 2DHGs we use the spin-$3/2$ operators, as heavy and light holes have effective spin $3/2$; $1/2$ from the electron spin and $1$ from the atomic orbital. Due to spin-orbit coupling, the total angular momentum $\bm J = \bm L + \bm S$ projected onto the crystal momentum is the sensible quantum number for this system.

We model the $z$ wave function using the Bastard wave function because it allows variation of both the well width and gate field independently, which gives it an advantage over a triangular well or harmonic oscillator model. We find that the orbital magneto-electric effect increases linearly with the gate field as shown in Fig.~\ref{Fig:Ge_OAM3D}. We also generally find that the effect is larger for wider quantum wells, as can be seen in Fig.~\ref{Fig:well_width}. However, since we only consider the ground states of the $z$-wavefunctions, there will likely be additional corrections with the inclusion of excited states, and excited states become more important for wider quantum wells. As such, we do not expect this relationship between the size of the OME and the well width to hold for any size quantum well. 

\section{Discussion}
Our main finding is that 2 dimensional hole gases with broken inversion symmetry exhibit a sizable in-plane orbital magneto-electric effect. The OME occurs due to coupling between heavy hole and light hole states with different centres of mass in the out-of-plane direction. As such, this OME is novel, as out-of-plane movement is often forbidden or ignored in 2D systems \cite{OM-PRB-2021-Xiaocong, OM-metal-PRB-2021-Xiaocong, OEE-NL-2018-Shuichi, osumi2021kinetic, PhysRevB.107.094106, cullen2025quantum}. We expect that there will be a significant OME in all $p$-type semiconductor 2DHGs with broken inversion symmetry. However, we expect Ge to be the best candidate for making orbitronic devices because of its high mobility and proximity to Si fabrication. The OME in both Ge and GaAs is enormous, we find that for realistic parameter values it is of the order $10^{12}$ $\hbar/$cm$^{2}$. Our results in this work, combined with our recent prediction of a massive OHE in Ge \cite{cullen2025ge}, should put Ge forward as a prime candidate for the realisation of efficient orbitronic devices. 

\begin{figure}[tbp]
\begin{center}
\includegraphics[trim=0cm 0cm 0cm 0cm, clip, width=\columnwidth]{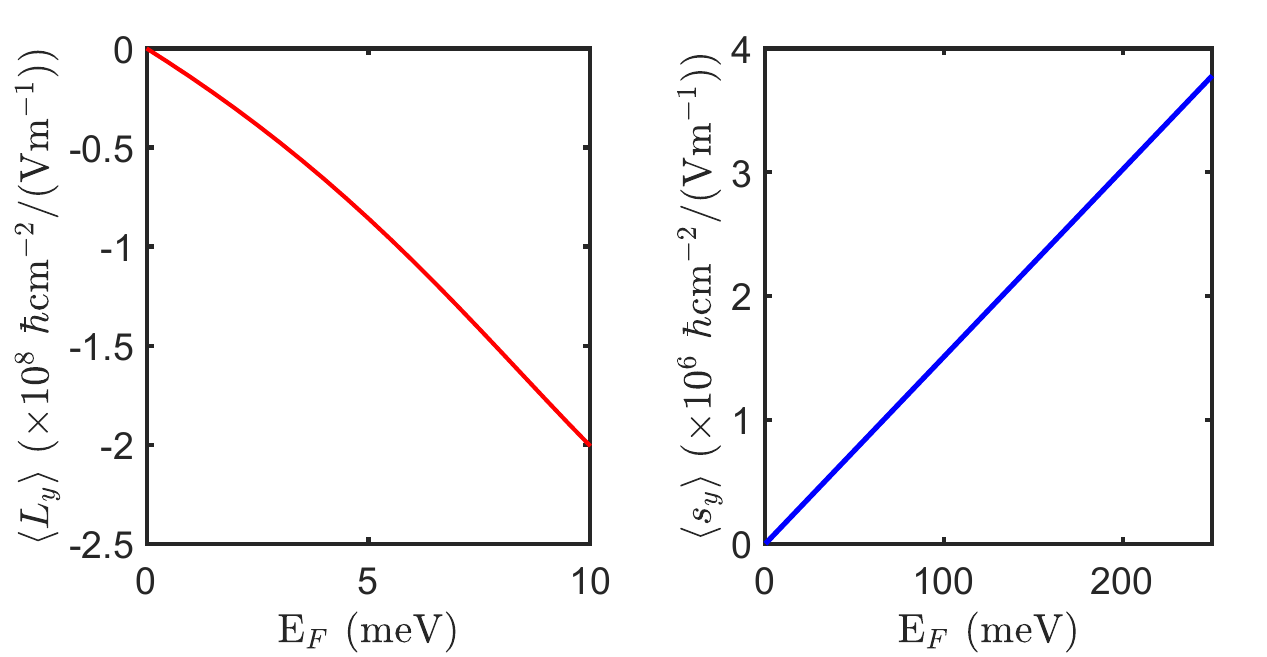}
\caption{\label{Fig:TivsGe}The (a) OAM density per unit field in a Ge 2DHG and, (b) spin density per unit field in TI surface states. The 2DHG has a confinement of 20 nm and a gate field of $F=10^5$ Vm$^{-1}$. {The TI surface state has Fermi velocity $v_F = 6.1\times10^5$ m/s. We choose typical relaxation time values of $100$ ps for the 2DHG and $1$ ps for the TI.}}
\end{center}
\end{figure}

We find the OME in Ge and GaAs to be $1-2$ orders of magnitude larger than our estimates of the Rashba-Edelstein effect in the surface states of topological insulators (TIs), materials known for their efficient charge-to-spin conversion \cite{SOT-TM-Rev}, as is demonstrated in Fig.~\ref{Fig:TivsGe}. This difference in magnitude is largely due to high mobilities in Ge and GaAs. Ge and GaAs 2DHG devices have demonstrated mobilities of the order order $10^6$ cm$^2/$Vs \cite{GaAsholemobility, stehouwer2023germanium, Borsoi2025nature}, indicating that their relaxation times are of the order $100$ ps, while TI surface state mobilities are generally of the order $10^3$-$10^4$ cm$^2/$Vs \cite{gehring2012growth, koirala2015record, shrestha2017extremely, wang2017observation}, corresponding to relaxation times of the order $0.1$-$1$ ps. {Our relaxation time estimates are determined by comparing current density calculations with experimentally measured mobilities}. 
{ The largest experimental mobilities referred to here were recorded at low temperatures $<1$ K, however given the enormous size of the OME we have calculated we still expect the OME in Ge to be significant at higher temperatures in which the mobility and scattering time are expected to be reduced.} 
It should be noted that for equivalent relaxation times, we find the REE in the TI surface states and the OME in 2DHGs to be of the same order of magnitude. 

{The novel OME calculated in this work occurs due to itinerant out-of-plane motion of holes. The out-of-plane motion considered here is itinerant as the matrix elements of the out-of-plane position operator are determined by the envelope wavefunctions which span many unit cells. For a system with an asymmetric confinement potential, the heavy hole and light hole wavefunctions will have different centres of mass in the confinement direction. The differing centres of mass allow for motion in the out-of-plane direction, which is usually forbidden in 2D systems, and the interband elements of the out-of-plane position and velocity operators are nonzero. When these matrix elements are multiplied by matrix elements of the in-plane position and velocity operators in the OAM operator $\bm L$, it results in a nonzero in-plane OAM in the hole eigenstates at each wavevector $\bm k$. Whereas in equilibrium the total OAM is exactly zero, as time reversal symmetry in preserved, out-of-equilibrium the Fermi surface shift by an applied electric field will generate a nonzero OAM similarly to how a spin density is generated in the Rashba-Edelstein effect.} 
{We expect an analogous OME mechanism to occur in multilayer 2D systems with broken inversion symmetry such as van der Waals heterostructures and some transition metal dichalcogenides.} 

For the in-plane components of the position operator that appear in the OAM operator the delocalised nature of Bloch electrons requires careful treatment \cite{cullen2025quantum}. This is because the intra-band elements of the position operator between Bloch wavefunctions are differential operators \cite{blount1962formalisms, Position-operator-Rev,resta1998positionop}. Due to this it is useful to introduce the effective displacement $\bm \Xi$ to calculate the expectation value of operators involving the position operator\cite{Hong-PSHE-PRB, liu2025quantumOHE, cullen2025quantum}. However, this problem does not affect the out-of-plane component of the position operator for the 2DHG system considered here, as holes are localised due to the confinement potential and all matrix elements of the $z$ component of the position operator can be straightforwardly computed. However, the localisation of holes in the confinement direction presents another problem to consider when computing the position operator, the choice of origin $z=0$, this affects the calculation of $(\bm r\times \bm v-\bm v\times\bm r)$ and can yield an OAM related to the magnetic field generated by a current. To avoid calculating such terms, we separated the matrix elements of the position operator in terms of the relative displacement of the heavy and light holes $\Delta z$ and the average position of the heavy and light holes $\Tilde{z}$. To avoid including terms that arise due to the choice of origin in our results, we set $\Tilde{z}=0$. However, it should be noted that $\Tilde{z}$ depends on the gate field, and as such our choice of origin also depends on the gate field.

The OAM density calculated in this work entirely comes from the band diagonal part of the nonequilibrium density matrix $n_{\bm E}$ and is linear in the relaxation time $\tau$ and dependent on extrinsic disorder. We have calculated the intrinsic disorder independent components of the OME along the lines of Ref.~\onlinecite{cullen2025quantum}, but found all components to be zero, this calculation is briefly covered in the supplementary material. As such, the size of the OME in 2DHGs is entirely dependent on how clean the sample and interface are. The intrinsic contribution to the OME occurs due to interband coherence induced by an applied electric field, as we recently showed this occurs in part via Zitterbewegung and the electrically induced dipole as well as the dipole dynamics in Bloch electrons \cite{cullen2025quantum}. The extrinsic mechanism that we calculate in this work occurs due to the following factors; the equilibrium eigenstates of the system have some degree orbital polarisation at each wave vector, however, the net orbital polarisation is zero due to the system having time reversal symmetry. When an external electric field is applied, the Fermi contour is shifted, proportionately to the relaxation time, and a net OAM density is induced in the system. The disorder in our calculation is treated very simply using a relaxation time approximation, there may be additional corrections to our results with a proper treatment of disorder. However, a full treatment of disorder including spin-orbit scattering effects is beyond the scope of this work, and will be addressed in a future publication.

The calculations of OAM density in this work have been done from the approach of the modern theory of orbital magnetisation, in which the expectation value of $m\bm r \times \bm v$ is calculated between Bloch states. The modern theory is one of two main approaches to the calculation of orbital effects. The other approach is the atom-centred approximation, which calculates the OAM purely from the OAM of atomic orbitals that make up each set of bands. {The OME calculated in this paper is due to itinerant motion over many unit cells and is not captured by the ACA.} For an ab initio model the atom-centred approximation should be contained within the modern theory, as the atom-centred approximation is derived from only including the local circulation part of the OAM\cite{OHE-PRB-2022-Manchon}.

Part of the immense interest in the generation of OAM is based on its use in orbital torques. In an orbital torque, an OAM density/current is generated in an orbitronic material, it is then converted to spin, which then exerts a torque on a magnetisation through the exchange interaction. One of the largest questions looming in the field of orbitronics is the mechanisms behind the orbital-to-spin conversion in orbital torques. From the perspective of the atom-centred approximation, orbital-to-spin conversion can be achieved through atomic spin-orbit coupling of the form $\alpha\bm L\cdot\bm s$, through which OAM can be converted to spin which is aligned parallel or anti-parallel to the OAM depending on the sign of $\alpha$ \cite{OT-PRR-2020-Hyun-Woo, CIAM-PRR-2020-Yuriy}. However, this mechanism still lacks rigorous theoretical justification; additionally, it does not include the Bloch electron OAM from the modern theory. This makes the accurate detection of the OME predicted in this work difficult, as 2DHGs occur at interfaces which are buried in the sample restricting optical detection, and as such the most sensible route is through the orbital torque. As discussed, orbital-to-spin conversion mechanisms within the modern theory are still largely unexplored, and the conversion efficiency in different materials is open question{, as such reasonable predictions of the conversion efficiency remain elusive}. 

{Indirect detection of the OME in 2DHGs through an orbital torque could potentially be achieved in a FM/Ge or FM/X/Ge heterostructure, where the X layer is used to assist with orbital-to-spin conversion. Alternatively, if an orbital density/current is injected into a Ge 2DHG via an adjacent material the inverse effect could be directly measured by the resulting current.} 
{The largest mobilities in Ge 2DHGs have been measured in Ge/SiGe heterostructures \cite{stehouwer2023germanium}, these devices have been engineered to have a very low impurity density. Engineering a device for orbitronic applications from Ge would likely have significant material challenges, as orbital torque devices would need to contain a FM layer, which could affect the quality of the Ge 2DHG through interface roughness, alloy disorder and lattice mismatch. We expect that such disorder would decrease the momentum relaxation time and in turn the magnitude of the OME.} 
{ For both proposed methods of detection of the OME, unless the system has been tuned such that the Fermi energy lies in the bulk gap, there will also be an almost indistinguishable signal coming from the OHE in the bulk, which has recently been shown to be large in Ge \cite{cullen2025ge}.}


In our calculations, we have used a number of approximations, mainly: The spherical approximation in the $4\times 4$ Luttinger Hamiltonian and, including only the ground state wavefunctions. First, for the main two materials considered GaAs and Ge, the spherical approximation works very well, and we expect corrections due to including the exact values of $\gamma_2$ and $\gamma_3$ to be minor. Lastly, while only including the ground state $z$ wavefunctions restricts the accuracy of our calculation for larger values of the Fermi energy, we expect that corrections due to excited states will not change the order of magnitude of the OME calculated, {further discussion of the effect of excited states can be found in the Supplement}. The sizable magnitude of the OME in Ge and GaAs that we have calculated is the main thrust of this work, since such inaccuracies resulting from the use of these approximations are secondary.

\section{Conclusions and Outlook}
We have shown that a sizable in-plane orbital magneto-electric effect is present in 2 dimensional hole gases with broken inversion symmetry. We showed that for realistic parameter values the electrically induced OAM density in Ge is of the order $10^{12}$ $\hbar/$cm$^{2}$, two orders of magnitude larger than the REE predicted in the TI surface states. As such, this work along with our recent prediction of a massive OHE in Ge, make Ge a prime candidate for building future orbitronic devices.

\section{Supplementary Material}
The supplementary material includes further details of the orbital magneto-electric effect calculations presented in the manuscript.

\section{Acknowledgements}
This work is supported by the Australian Research Council Discovery Project DP2401062.


%

\end{document}